\documentstyle[aps,prc,preprint] {revtex}

\begin{document}

\title {Generalized Entropy and Temperature in Nuclear
Multifragmentation }

\author{   A. Atalmi, M. Baldo, G. F. Burgio and A. Rapisarda }

\address{ Istituto Nazionale 
di Fisica Nucleare, Sezione di Catania
and Dipartimento di Fisica, \\   Universit\'a di Catania,
Corso Italia 57, I-95129 Catania, Italy }

\date{\today}
\maketitle

\begin{abstract} 

In the framework of a 2D Vlasov model, we study the time evolution of 
the "coarse-grained" Generalized Entropy (GE)  in a  nuclear system 
which undergoes a multifragmentation (MF) phase transition.
We investigate the GE both  for  the gas and 
the fragments (surface and  bulk part respectively). 
We find that the formation of the surface causes  
the growth of the GE during the process of fragmentation.
This quantity then characterizes the MF and confirms the crucial role 
of deterministic chaos in filling the new available phase-space:
at variance with the exact time evolution, no entropy change is 
found when the linear response is applied.
Numerical simulations were used also to extract information about 
final temperatures of the fragments.
From a fitting of the momentum distribution with a Fermi-Dirac  
function we extract the temperature of the fragments 
at the end of the process.  
We calculate also the gas temperature by averaging over the available phase 
space. The latter  is a few times
larger than the former, indicating  a gas not in equilibrium.  
Though the model is very schematic, this fact seems to 
be very general and could explain 
the discrepancy found in experimental data  when using the slope
of light particles spectra instead  of   the double ratio of isotope yields
method in order to extract the nuclear caloric curve.

\end{abstract}



Multifragmentation phase transition is one 
of the  most interesting recent discoveries in nuclear physics
\cite{finn,sie}. 
In the last years an  intense effort 
has been put forward both experimentally 
\cite{gsi,eos1,eos2,indra,mastinu,fopi,aci}
and theoretically 
\cite{aci,bo,gross,buu,bert,kap,campi,bauer,cmd1,bur92,bbr1,bbr2,jac,lin,cmd} 
in order to understand this phenomenon.
Several scenarios have been proposed for the
onset mechanism and for the dynamics underlying the Nuclear
Multifragmentation (MF). 
In the statistical model \cite{bo,gross}
one assumes that the energetically available phase space dominates
the reaction dynamics. This implies that the set of multifragmentation
events fills, all together, the phase space in an almost uniform fashion
and therefore a quasi-static statistical description is possible.
The model appears quite successful in describing some of the
phenomenological features observed in many MF 
experiments, including the multiplicity and mass distributions
at different asymmetries, as well as the more recently
discussed ``caloric curve", which some groups claim to have extracted 
from the observational data \cite{gsi,eos1,eos2,indra}.
The model, however, does not provide a mechanism for the formation
of the fragments along the dynamical evolution of the reaction,
and for the filling of the available phase space (once the
whole ensemble of multifragmentation events is considered). The latter
feature implies a large (maximal) production of entropy, which is
indeed proportional to the logarithm of the phase space volume occupied 
by the considered ensemble of events \cite{gross}. A 
natural source of entropy
is the process of collisions between particles, as, for instance,
in the BUU model of nuclear reactions \cite{buu,bert,kap} 
or in molecular dynamics
simulations (see for example ref. \cite{cmd1}). 
It has to be kept in mind,
however, that if the system is chaotic \cite{bbr1,bbr2,jac,cmd}, 
large fluctuations
from one event to another must be expected, with a possible filling
of the phase space by the ensemble of events, independently from
the details of the  microscopic mechanism.\par

In order to contribute to the clarification of the previous points, 
we discuss in the present paper new calculations of 
nuclear multifragmentation in 
a schematic and simplified
2D Vlasov model already well known in the 
literature \cite{bur92,bbr1,bbr2,jac}.  
We start the system inside the mechanically unstable spinodal 
region and follow the birth and growth of fragments induced by a 
very small initial random noise. The latter simulates the missing initial 
dynamics and drives the system outside the spinodal region. In order
to characterize this process 
we use the "coarse-grained" Generalized Entropy
(GE) which is the non-equilibrium extension of the thermodynamical
entropy \cite{boltz}. 
GE has already been used in the past both theoretically \cite{kap} 
and experimentally \cite{fopi} in order to investigate nuclear MF.
Usually the increase of GE is believed to be due to two-body collisions.
We find that GE grows even in a Vlasov approach if initialized inside
the spinodal region. More precisely we find that GE grows as the surface 
of fragments is formed.
This growth reflects the crucial role of chaoticity in MF confirming what 
has already been found in previous studies \cite{bbr1,bbr2,jac,cmd}. 
Chaotic motion is the main mechanism responsible for dynamical filling 
of the available  equilibrium phase space. 
This result seems to be very general
in phase transitions as recently confirmed in several investigations 
\cite{cmd,nayak,posch,pettini,latora}. 

Recently a very interesting link 
between the GE and the Kolmogorov-Sinai entropy (KS entropy) (the sum
of the positive Lyapunov exponents) \cite{arnold,schuster} has been found for 
chaotic maps \cite{baranger}.
 Our results are consistent with  this finding
in that no entropy growth is observed when the linear response is applied
(regular evolution) \cite{jac,lin}.

In the present  paper we extract also  
information about the temperature of the fragments. 
We find also a gas of particles which is not in 
equilibrium with the fragments. The gas has a temperature which is
a factor of 2-3 greater than the fragments' one. 
These non-equilibrated 
particles in a realistic situations will likely be emitted  in a first stage
of the process and do not carry information on the equilibrated 
component of the system which undergoes MF  \cite{eos2}. 
Though the model is very schematic
this finding can be very general and explain the puzzling discrepancy 
observed experimentally between  the caloric curve extracted from isotope
yields \cite{albergo} and the one from the slopes of light particle spectra 
\cite{eos2,indra}.

This paper is organized as follows. In section I we remind the reader the 
details of the model. In section II we define the "coarse-grained" GE 
and discuss its connection to chaos and thermodynamical entropy.
Numerical results are presented in section III and conclusions are drawn 
in section IV. 
We report in the appendix the method we used to extract the temperature.

\section{ The Model}

Our theoretical framework is based on the solution of the Vlasov 
equation for the one-body density {\it f} in phase space

\begin{equation}
{\partial f({\bf r}, {\bf p }, t) \over \partial t}= \{h[f], f\}
\label{eq:vla}
\end{equation}

Here $\{ .,. \}$ is the Poisson bracket, so that
eq.(\ref{eq:vla}) represents the collisionless propagation of {\it f} in the
self-consistent one-body field described by the effective 
single-particle hamiltonian
$h({\bf r}, {\bf p}) = p^2/2m + U({\bf r}, {\bf p})$, being U the
self-consistent mean field.  Over the past decade the Vlasov equation
has been widely used in nuclear mean-field theory
for describing many aspects of intermediate energy heavy ion collisions.
In particular it has been extended by incorporating a Pauli blocked
collision term leading to a Boltzmann-like dynamical description,
denoted the BUU \footnote{Various names have been adopted in literature,
{\it i.e.} Boltzmann-Uehling-Uhlenbeck,   
Landau-Vlasov or Vlasov-Uehling-Uhlenbeck equation.} model\cite{buu,bert}.
In the present paper no collision term has been considered, since we
have already seen \cite{bbr1,bbr2} that it is not very important inside
the spinodal region.
 
In the present work the Vlasov equation has been solved
numerically in a two-dimensional lattice using
the same code of ref.\cite{bur92}, as already done in refs. \cite{bbr1,bbr2}.
We have studied a fermion gas situated in a two-dimensional 
periodic box whose size is kept constant during
the evolution. The box sidelengths are equal to $L_x = 51~fm$ and
$L_y = 15~fm$. We divided the single particle phase space into several
small cells. We employed in momentum space 51x51 small cells of
size $\Delta p_x = \Delta p_y = 40~ MeV/c$, while in coordinate
space $\Delta x = 0.3333~fm$ and $\Delta y = 15~fm$. 
The initial local momentum distribution was assumed to be the one of
a  Fermi gas at a fixed temperature $T$. 
For the effective one-body field we employed a simplified Skyrme
interaction   $U[\rho] = A~(\bar \rho(x) /
\rho_0) + B~(\bar \rho(x) / \rho_0)^2$. 
The saturation density in two dimensions 
$\rho_0 = ~0.55~fm^{-2}$  corresponds to the usual
three-dimensional  Fermi momentum of $P_F =~260~MeV/c$.
Furthermore $\bar\rho(x)$ is the local average of the density 
with respect to the transverse direction y, smeared in the  
$x$-direction with a gaussian of width $\mu = 0.61~fm$, in order
to give a finite range to the interaction. The parameters of the
force $A$ and $B$ have been chosen in order to reproduce
correctly the binding energy of nuclear matter at zero temperature,
and this gives $A = ~-100.3~MeV$ and $B = ~48~MeV$.  
Then a complete dynamical evolution is performed by subdividing the
total time in small time steps, each equal to $\Delta t = 0.5~fm/c$.
The algorithm is deterministic, i.e. no noise of the Langevin-type has been 
used, and the total energy is conserved with a good accuracy.
For more details concerning the mean field propagation on the
lattice, the reader is referred to ref.\cite{bur92}.

\section{Thermodynamical Entropy, Generalized Entropy and chaos} 

In this paper we study the 
amount of entropy generated during the time of fragment formation. 
For this purpose a few considerations are necessary in order to
introduce the notations and some quantities which will be
essential in the numerical analysis.
\par
Let us start by summarizing some basic concepts of statistical 
mechanics. Consider a hamiltonian system with $n$ degrees of
freedom. The thermodynamical entropy $S_0$ is obviously
defined only at thermodynamical equilibrium and is directly related
to the number $N$ of microstates which correspond to a given
macroscopic state. If we put equal to 1 the volume occupied in
phase space by each one of the microstates (e.g. we put $\hbar = 1$ in the 
quantal case), $N$ coincides also with the available microscopic 
phase space $\Omega$ (the volume of a $2n$-dimensional manifold 
in the classical case). Historically this connection was 
introduced by Boltzmann and provides  the statistical basis 
of thermodynamics \cite{boltz,mesta,penrose}. In general, 
$S_0$ is just proportional to 
the logarithm of the number $N$ of the possible microscopic states 
in the considered macroscopic conditions. The latter defines
 the ``ensemble" 
of microscopic states which correspond to the given macroscopic 
thermodynamical state. Therefore, in the microcanonical ensemble
the entropy is just the logarithm of the microscopic
available phase space $\Omega(E)$ at a given total energy $E$ 
$$
   S_0 = \ln N(E) \equiv \ln \Omega (E) = -\ln P(E)
$$
\noindent
where $P(E) = N(E)^{-1} \equiv \Omega^{-1}$ is the probability 
of each one of the microstates, which are assumed to have all the same 
{\it a priori} probability. \par
Once the entropy is given in the microcanonical ensemble, 
all the other thermodynamical quantities can 
be extracted, or other ensembles can be introduced by means of a suitable 
Legendre transform. \par
Let a real macroscopic system be prepared in a generic microstate.
Then, it is possible to describe the system according to standard statistical
thermodynamics only if it is at least ergodic, namely the system in
its temporal evolution will sample all the microstates (or an
overwhelming majority of them) and with the same frequency. Since the energy
is conserved, the asymptotic temporal average of the frequency 
will then coincide indeed with $P(E)$. However, it is essential to
recognize that, for a system of identical particles, sets of very large
number of equivalent configurations exist, namely the configurations
 which differ only
by the 
rearrangement
of particles among equivalent microstates. Among these sets, the 
one which contains the largest number of configurations will be the
dominant one, since it will be the one with the highest occurrence frequency 
and statistical weight. At equilibrium the system will move mainly within
this set of configurations (at least for large systems, for which
fluctuations can be neglected). For instance, for a system of weakly
interacting
particles the dominant set is nothing but the one characterized
by the canonical (Boltzmann) single particle occupation numbers.
Of course, the time of approach to equilibrium is in general 
strongly dependent on the initial conditions and on how fast the
system explores the available phase space.\par
In the physical conditions considered here, we are however more interested
to what extent a set of initial conditions, taken all together,
explore the available phase space during the evolution of the system.
It is in fact the whole set of MF events which has
to be considered, since it is the ensemble over which  the relevant 
physical quantities have to be calculated. It can be also useful
to consider different sub-sets of MF events, in order to study the 
fluctuations which are present within the whole MF set of events. 
In any case, this procedure is equivalent to the 
introduction of a Gibbs ensemble, namely a set of copies of
the system at the initial time with slightly different initial conditions. 
Correspondingly, the ability of the set 
of initial conditions
to reach equilibrium
is usually referred to as the ``mixing" property \cite{penrose}.
Mixing means that a generic set of nearby initial states, 
occupying a small initial volume $\omega_o$ in phase space, away
from equilibrium, will spread 
rapidly throughout the available phase space, eventually dominated
by the equilibrium configurations.
It can be proved that mixing implies ergodicity \cite{arnold,schuster}.

This spreading, therefore,
describes the evolution towards equilibrium of the system when
initially prepared in a generic state inside the volume $\omega_0$.
If one wants to describe quantitatively the approach to equilibrium,
one can introduce a generalized and time dependent entropy
just as the logarithm of the volume $\omega(t)$ occupied at time
$t$ by the states originating from the set of initial
conditions inside the volume $\omega_0$ at $t = 0$. In other words,
one can follow the evolution of the initial volume $\omega_0$
under the automorphism defined by the equations of motion. 
\par
As it is well known \cite{penrose}, this approach 
faces a well defined difficulty
in the case of hamiltonian systems : due to Liouville theorem,
the volume in phase space is conserved, and according to the
above mentioned definition no entropy could be produced.
\par
However, the filling of phase space for a
hamiltonian system which displays chaoticity can be very intricate and
``filamentary". In this case a ``coarse grained" description appears quite
natural, and the introduction of a GE function
is possible and useful \cite{penrose}. 
We illustrate this point with an example, taken
from the literature on dynamical systems \cite{schuster}. 
Let us consider the standard
map, which has the property of conserving the phase space area, and
therefore it is quite analogous to a hamiltonian system. Actually
it can be recasted into a hamiltonian form \cite{schuster}.  If we consider a
set of nearby initial conditions, confined inside 
the square depicted in fig.1 (top panels), 
the map
$$
      y_{n+1} = y_{n}  + {K\over{2\pi}} sin(2\pi x_n)  ~~~~~~~mod~1
$$
$$
      x_{n+1} = x_{n}  +  y_{n+1}  ~~~~~~~~~~~~~~~~~~~mod~1
$$
\par\noindent
(K is the coupling term) will 
spread the initial occupied volume into a very complex and scattered 
structure, as the number of iterations {\it n} increases. 
We report the time evolution for two values of the coupling, i.e. K=0.05
and K=2. In the first case one has an almost regular evolution, while in the 
second case a chaotic dynamics sets in. 
The more chaotic is the map, the larger is the area of phase space where the 
ensemble of points is scattered.
In the case K=2, after only a few iterations the equilibrium 
phase space is filled almost 
uniformly by a net of fine structures. 
One can compare this mixing property of such a chaotic map with the diffusion 
of a drop of ink inside a glass of water. 
On the other hand the quasi-regular case (K=0.05) would correspond to a 
drop of mercury. 

In order to study the filling of phase space
one can then divide the available phase space
into a set of ``coarse grained" cells, and take the fraction $p_i$
of the filled area inside each cell $i$. Equivalently, if one
considers a set of representative points inside the initial square,
one can define $p_i$ as the fraction of points inside the 
cell $i$ at a given time $t$.
The coarse grained GE can be then defined by

\begin{equation}
 S \, = \, - \sum_i p_i \ln p_i   \label{eq:entr}
\end{equation}
\par\noindent

In a ``fine grained" description, the cell size
is considered arbitrarily small, thus one 
has $p_i \, = 0\,$ or  $p_i \, = 1\,$,
and the entropy is identically zero. At variance, 
in a coarse grained description,
$0 \, < p_i \, < 1$, so that  $S$ is different from zero and it
increases as the phase space filling increases. 
If the spread
is uniform in phase space, one obtains  the standard thermodynamical 
equilibrium entropy, with
the microstates specified by the N cells of the  lattice.
In fact, in this case $p_i=1/N$ and $S=\ln N= S_0$.

\par
The increase of entropy in the considered example
as a function of the number of iterations is reported in fig.1 (bottom panel)
for the two cases. Please
notice that this definition of GE  can be introduced
also for non-mixing systems, for which the available phase space
is not necessarily full or uniformly filled in the large time 
limit. Therefore the GE is a very useful tool to measure how big is 
the available equilibrium phase space once a proper grid is given
to measure it. 
In general the definition of GE is not unique and depends on the grid size
\cite{penrose}.

It has to be stressed
that this definition of GE  is physically different from the
one adopted in the theory of dynamical systems by Kolmogorov \cite{arnold}. 
The KS entropy  is the 
sum of the positive Lyapunov exponents \cite{schuster}.
 Recently a very interesting connection between the GE and the KS entropy 
has been found for chaotic maps \cite{baranger}: the latter corresponds
to the slope of the 
linear growth rate of the GE. Though this result is not a theorem and  
at the moment  it is not clear 
if it can be simply extrapolated to 
systems with many degrees of freedom as 
the one we have investigated, a growth of GE is a strong hint of 
chaoticity. In the following we will discuss numerical results which 
support this conjecture.

A coarse grained description is introduced in the Vlasov
equation (\ref{eq:vla}), once it is written for a discrete grid. 
Through the GE we can in fact investigate the filling of phase
space in MF.
 
In this case the differential equation (\ref{eq:vla}) becomes a set of 
finite difference equations
\begin{equation}
{\partial f( {\bf r} , {\bf p} ) \over \partial t} + 
 {\bf v}({\bf p}) \cdot {\bf \Delta}^r f + 
 {\bf F}({\bf r}) \cdot {\bf \Delta}^p f  = 0  \label{eq:vladis}
\end{equation}
\noindent
where (the index $i$ labels the vector components)
\begin{equation}
{\bf \Delta}_i^r f = {f( {\bf r} + {\bf n}_i , {\bf p} ) - 
                      f( {\bf r} - {\bf n}_i , {\bf p} ) \over 2 \Delta r_i }
\end{equation}
\begin{equation}
{\bf \Delta}_i^p f = {f( {\bf r}  , {\bf p} + {\bf m}_i ) - 
                      f( {\bf r}  , {\bf p} - {\bf m}_i ) \over 2 \Delta p_i }
\end{equation}
\noindent
{\bf v}({\bf p}) and {\bf F}({\bf r}) are respectively the velocity and the 
force.
In the last equations ${\bf n}_i$ and ${\bf m}_i$ are the vectors connecting
the centers of close neighbour cells in the direction $i$. The
non-zero components of these lattice vectors are the
previously introduced lattice cells size $\Delta r_i$ and
$\Delta p_i$ respectively. It is easily verified that the particles number
and the total energy are conserved also in this discretized form
of the Vlasov equation.\par
The GE can be then defined as
\begin{eqnarray}
 S = &{- g\over N_h}\sum_{{\bf r},{\bf p}} [ f'( {\bf r}  , {\bf p} )
 \ln f'( {\bf r}  , {\bf p} )  + &   \nonumber \\
  & ( 1 - f'( {\bf r}  , {\bf p} )) 
\ln ( 1 - f'( {\bf r}  , {\bf p} ) ) ] \label{eq:entropy}
\end{eqnarray}
\noindent
where $f'=f h^2$, 
$g$ is the spin-isospin degeneracy factor (g=4 in our case)
and $N_h =h^2 / \Delta \Omega$ is the number of cells contained
inside a volume of size $h^2$,
$\Delta\Omega$ being the cell volume
(this normalization ensures that the 
number of quantal states is correctly counted).
This definition goes into the usual (coarse grained) 
thermodynamical entropy in the case of thermodynamical equilibrium.

The variation of entropy, according to the finite difference equation of
motion (\ref{eq:vladis}), is given by
\begin{eqnarray}
{\partial S \over \partial t} =  {\Delta\Omega ~g\over N_h} 
   \sum_{{\bf r},{\bf p}}
 \sum_{ij} [ {v_i\over 2 \Delta r_i} f'( {\bf r} + {\bf n}_i , {\bf p} )
 \ln { f'( {\bf r} , {\bf p} ) \over
 f'( {\bf r} + {\bf n}_i , {\bf p} ) } + ~~~~~ & \nonumber \\
 {F_j \over \Delta p_j} f'( {\bf r} , {\bf p} + {\bf m}_j )
 \ln { f'( {\bf r} , {\bf p} ) \over
 f'( {\bf r} , {\bf p} + {\bf m}_j ) } + ~~~~~ & \nonumber \\
 {v_i\over 2 \Delta r_i} (1-f'( {\bf r} + {\bf n}_i , {\bf p} ))
 \ln { (1-f'( {\bf r} , {\bf p} )) \over
 (1-f'( {\bf r} + {\bf n}_i , {\bf p} )) } + ~~~~~ & \nonumber \\
 {F_j \over \Delta p_j} (1-f'( {\bf r} , {\bf p} + {\bf m}_j ))
 \ln { (1-f'( {\bf r} , {\bf p} )) \over
 (1-f'( {\bf r} , {\bf p} + {\bf m}_j )) } ] ~~~~~
\end{eqnarray}
\noindent
which is different from zero whenever the distribution $f$ is not uniform
and $\Delta \Omega \ne 0$.
In particular spontaneous symmetry 
breaking can occur as in the case of chaotic dynamics. 
Notice that 
${\partial S \over \partial t} = 0$,
,  if $f'=1/2$, as it happens e.g.
at the Fermi energy for a Fermi distribution. 
This is exactly what happens when the linear 
response is applied 
to a uniform system at a given temperature.
We will discuss
later this important point.

Similarly we define the corresponding GE  per particle
$\sigma $ as
\begin{equation}
\sigma = {S \over \sum_{{\bf r},{\bf p}}  f( {\bf r} , {\bf p} ) } = 
{S \over A}
\label{eq:part_entropy} 
\end{equation}
being A the total particle number. \par
It has to be noticed that in the usual continuous Boltzmann equation 
the source of entropy is the collision term, which is treated within
the assumption of molecular chaos. This assumption, as it is well known, 
is an {\it ad hoc} hypothesis, which introduces irreversibility
in the otherwise reversible evolution of the classical system. Here we
are exploring a complementary source of irreversibility, namely
the filling of phase space due to the strong chaoticity of the time
evolution of the system at the mean field level. 
Notice, in fact, that the entropy of eq.(\ref{eq:entropy}) is a single 
particle entropy,
while, in general, one should consider a $N$-body entropy, for which
the distribution in the full 
$2N$-dimensional phase space should be involved.
Anyhow, the addition of the collision term could 
further increase the GE production rate also of the single particle
entropy.
It should be noticed that the 
GE for the Landau-Vlasov equation has already been used in heavy-ion
collisions, see for example refs.\cite{kap}. 

\section{Multifragmentation: results and discussion}

\subsection{Entropy production: exact numerical results}

In order to investigate the process of fragment formation, we 
analyzed the behavior of nuclear matter in the spinodal 
zone of the phase diagram, where uniform matter is unstable with 
respect to density fluctuations. 
For this purpose we chose the average nuclear matter initial density 
smaller than $2/3~ \rho_0$ 
and the temperature $T = 3~MeV$, i.e. nuclear matter 
is prepared well inside the unstable region. 
Since in our simulations we neglected the dynamical evolution which
drives the system inside the spinodal zone, we imposed a very small and 
uniform white noise on the initial average density profile. This 
random initialization will mimic the initial missing dynamics, perturbing 
the unstable equilibrium and 
forcing the system to relax towards a more stable
configuration.
Within this scheme, the time
evolution of a typical single trajectory, as well as the one of a 
bunch of trajectories, has been extensively analyzed in literature 
\cite{bur92,bbr1,bbr2}. In particular, the appearance of deterministic
chaos after an initial linear evolution plays a crucial role in the
process of fragment 
formation \cite{bbr1,bbr2}. The details of the interaction
could slightly modify the 
evolution time as found in ref.\cite{jac}, but
not the character of the dynamics.\par
With reference to the discussion 
in the previous section, we have to stress
that, with the above mentioned initial conditions inside the spinodal
region, the equilibrium configuration corresponds to non-homogeneous
matter, since fragments are spontaneously formed during the evolution
due to the instability of the system. The equilibrium configuration should
correspond to a well defined distribution of fragments in thermal equilibrium
among each other and with a vapour, but, due to the finite size of the 
system, large fluctuations in the fragment distribution occur
from one event to another. For this reason we do not consider the
whole set of possible outcomes all together, but we prefer to
initialize the system according to the above mentioned prescriptions,
which correspond to include and average over a
 sub-set of MF processes for each event
(i.e. a single computer run). This sub-set is specified by the initial 
distribution in phase space as described by the occupation 
probability $f({\bf r}, {\bf p })$ which specifies the average number of 
particles $f \Delta \Omega$ in each phase space cell and therefore gives a 
classical coarse grained description of the initial condition.

A typical single event is shown in fig.2, where we plot the time evolution 
of the density $\bar \rho(x)$. The initial average density 
is taken equal to $0.55 \rho_0$. We notice that the small random 
noise on the initially uniform density profile (please notice the different 
scale used in the top panel of fig.2) is rapidly amplified by
the action of the effective one-body field, thus leading towards 
fragment formation. An analysis of the Fourier spectra of the excited 
modes has shown a strong coupling with a sensitive dependence on the 
initial conditions \cite{bbr2}.

Following the definition of the previous section we show in fig.3 
(panels (a) and (c)) the time variation of the 
fraction of occupied phase space cells $N/N_{tot}$
for MF events started at initial density 
$\bar \rho / \rho_0=0.55$ and the corresponding coarse-grained GE's 
(panels (b) and (d)).  
In the figure we display the simulation for different widths of the gaussian 
$\mu$ used to smooth the density in the x-direction 
and different sizes of the cells. In panels (a) and (c) we observe an 
increase of the fraction of occupied phase space cells $N/N_{tot}$ 
at the time corresponding to the fragment formation, being the latter
dependent on the gaussian width $\mu$ \cite{jac},
i.e. on the range of the interaction and on the size of the grid. More 
precisely we get a slower evolution when using a larger $\mu$ 
\cite{jac} and a greater increase when increasing the size of the cell. 
We notice that the filling of phase space shows the same time evolution
of the GE, which is shown in panels (b) and (d).
Therefore MF is  strongly characterized  by a substantial increase in 
the GE which 
reflects, as in the example of the standard map previously discussed, the 
dynamical filling of phase space due to chaotic motion. The GE is a 
powerful tool to measure the volume of the occupied phase space,
though in a relative way due to the grid dependence.
All the calculations that will be shown in the following have been performed 
using $\mu=0.61 ~fm$ and $\Delta x= 0.3333 ~fm$. 

It can be very instructive to calculate the different contributions to the
entropy $\sigma $ coming from the surface and the volume of the formed
fragments. For this we have to define exactly surface and volume of
the two-dimensional fragments formed during the dynamical evolution.

Let us define the {\it surface} of the fragments as the ensemble of coordinate 
space cells where the density $\bar \rho (x)$ takes values between the  
limits $\tilde \rho_1$ e $\tilde \rho_2$ defined by 

\begin{equation}
\tilde \rho_1 = \frac{1}{10}~ \rho_{max}  
\label{eq:rho_inf} 
\end{equation}

\begin{equation}
\tilde \rho_2 = \frac{9}{10}~  \rho_{max} 
\label{eq:rho_sup} 
\end{equation}

being $~\rho_{max}$ the highest value of the density $\bar \rho (x)$. 

Analogously we define the {\it bulk} of the fragments as the ensemble of 
coordinate space cells where the density $\bar \rho (x,y)$ is larger than
$\tilde \rho_2$ of eq.(\ref{eq:rho_sup}). For completeness we also
characterize the {\it gas} component as the ensemble of space   
cells where the density is smaller than $\tilde \rho_1$ of 
eq.(\ref{eq:rho_inf}). Briefly we can summarize by writing

\begin{equation}
\tilde \rho_1 < \bar \rho(x) < \tilde \rho_2 ~~~~~~ surface
\label{eq:surf} 
\end{equation}

\begin{equation}
\bar \rho(x) \geq \tilde \rho_2 ~~~~~~~ bulk
\label{eq:bulk} 
\end{equation}

\begin{equation}
\bar\rho(x) \leq \tilde \rho_1 ~~~~~~ gas
\label{eq:gas} 
\end{equation}

We have checked that this particular choice of the upper and lower 
density limits produces numerically robust results.

Once the fragment has been defined, we can easily calculate the relative 
contributions to the entropy per particle $\sigma$ coming from the surface
and the bulk of the fragments, and from the gas. 
For this purpose
we denote with $S_i$ the GE of the component $i$ and $A_i$ 
the corresponding number of particles. We define the entropy per particle of 
the component $i$ by

\begin{equation}
\sigma_i = \frac{S_i}{A_i} \label{eq:sigma_i}
\end{equation}

It can be easily checked that the total entropy per particle $\sigma$ can 
be expressed as a weighted sum of $\sigma_i$ over all components,
{\it i.e.}

\begin{equation}
\sigma = \frac{S}{A} = \frac{1}{A} \sum_i S_i = 
\sum_i (\frac{A_i}{A}) \sigma_i \label{eq:sigma}
\end{equation}

Therefore each component contributes to $\sigma$
with the following amounts 

\begin{equation}
\sigma_{surf} = \frac{-g}{A_{surf}~N_h}
\sum_{\tilde \rho_1 < \bar \rho(x) < \tilde \rho_2}
[f'~ ln f' + (1 - f')~ ln (1 - f')] \label{eq:sigma_surf}
\end{equation}

\begin{equation}
\sigma_{bulk} = \frac{-g}{A_{bulk}~N_h} 
\sum_{\bar \rho(x)  \geq \tilde \rho_2}
[f'~ ln f' + (1 - f')~ ln (1 - f')] \label{eq:sigma_bulk}
\end{equation}

\begin{equation}
\sigma_{gas} = \frac{-g}{A_{gas}~N_h}
\sum_{\bar \rho(x) \leq \tilde \rho_1}
[f'~ ln f' + (1 - f')~ ln (1 - f')] \label{eq:sigma_gas}
\end{equation}

The onset of MF can be characterized by the time
variation of the fraction $\alpha_i$ of coordinate space cells occupied by 
the gas, the bulk and the surface \footnote 
{The reader should notice that $\alpha$ is defined in ``coordinate space''
and is different from the previously 
defined $N/N_{tot}$, being the latter the number
of occupied ``phase space'' cells.}. For them the following relation holds

\begin{equation}
\alpha_{surf} + \alpha_{bulk} + \alpha_{gas} = 1 
\label{eq:alfa}
\end{equation}

This is clearly shown in fig.4, panels (a) and (c). In particular 
we display in panel (a) the time variation of the fraction of space cells
belonging to the bulk (circles) and the surface (dashed line),
whereas in panel (c) we show the variation of the fraction of space cells 
belonging to the gas (dotted line). Those results concern a system
initialized at density $\bar \rho = 0.55~\rho_0$. 
We clearly observe their sudden change at the time when fragments
form. In particular, the gaseous and the surface parts are zero before 
fragments form, whereas the bulk fraction is dominant. 
During fragment formation, their relative contribution change, 
the bulk contribution decreases while the gas and the surface grow 
until fragments are completely formed. At this time, which we define
as the fragmentation time $\tau_{frag}$, all 
the $\alpha's$ reach a plateau, apart from some small fluctuations.
We have checked that this behavior is quite general and does not depend
on the initial density, as can be seen in panels (a) and (c) of fig.5,
where the same results are plotted for an initial density 
$\bar \rho = 0.3~\rho_0$. 

At the fragmentation time an appreciable increase of 
the entropy per particle $\sigma$ is observed, see fig.4(b). 
There we display the time variation of the total entropy 
per particle $\sigma$ (solid line), the bulk (circles) and 
the surface entropy (dashed line) for a system 
initially prepared at an average density $\bar \rho = 0.55~\rho_0$. 
At fragmentation time $\tau_{frag} \simeq 160~fm/c$
the increase of the entropy per particle is about $\Delta\sigma \sim 0.4$
and is independent on the initial average density which, on the other 
hand, determines the fragmentation time. This is clearly shown in panel (b)
of fig.5, where the case for a system with initial density 
$\bar \rho = 0.3~\rho_0$ is displayed. Here the fragmentation time is shorter
than the previous case, $\tau_{frag} \simeq 80~fm/c$. 

Please note also the relative contributions of the surface and the bulk.
The surface entropy (dashed line) shows a strong 
increase from zero to 0.9. When the surface forms, the surface
entropy is practically equal to the total entropy per particle,
$\sigma_{surf} \sim \sigma$. We have checked that this relation actually 
holds for all initial average densities, see fig.5(b). Therefore
the main mechanism of entropy production appears to be the formation
of surfaces region, which in the final stage include a substantial fraction 
of the total number of particles.

On the contrary, the bulk entropy $\sigma_{bulk}$ (circles in
panel (b) of figs.4 and 5) does not change appreciably during the time of
fragmentation and keeps close to the initial entropy per particle
$\sigma$. 

As far as the gas is concerned, we notice that 
the process of fragmentation is followed by the formation of 
a very excited and rarefied gas. 
As one can see from figs. 4(d) and 5(d), the gas entropy per particle  
$\sigma_{gas}$ takes values about a factor of three bigger than the 
surface and bulk entropies, although large fluctuations
show up during the fragmentation. The gas component is, however,
quite small and gives an almost negligible contribution to the
total entropy $S$ and to the total entropy per particle $\sigma$.
Evidences of a larger entropy per particle of the gas component
has been found in the analysis of experimental data \cite{kap},
and the values we find in our simulations seem to be typical
for MF reactions \cite{kap,fopi}.
The gas temperature has been found
to be very high and will be discussed in the following \cite{fopi}.

\subsection{Entropy production: exact numerical results vs. linear response}

In this subsection we compare the 
exact numerical evolution with 
the one obtained using linear response \cite{lin} starting from the 
same initial condition. In 
agreement with ref. \cite{jac},
we find that in the initial stage of the time evolution the linear response
is a good approximation to  the exact one.
In the case reported in fig.6 for the initial density
$\rho=0.4 \rho_0$ (this initial density has been chosen in order to facilitate 
the comparison with ref.\cite{jac})
we see that up to 40-50 fm/c the two profiles are  similar,
but they differ drastically as time goes on. Moreover the linear response
violates strongly energy and number of particles conservation after 70-80
fm/c. At that time the density profile becomes also negative.
But apart from this considerations which, though qualitatively similar, 
depend  on the initial density
and the folding used for the response, the most important difference between
the two time evolutions is that linear response does not produce any growth of
GE at variance with the exact result (see fig.6(f)). This fact is easily 
understood considering what already noticed     
at the end of section II. The variation of entropy with time is zero if
the occupation number $f'$ changes around $1/2$, i.e. very close to the 
Fermi energy. This is what happens for the linear response which is not able 
to thermalize the system.
Therefore this finding 
confirms the crucial role of nonlinearity and  chaos, giving at the
same time a strong support to the conjecture advanced 
in ref.\cite{baranger} for chaotic maps. Unfortunately it is very 
difficult in our case to calculate all the Lyapunov spectrum and verify 
that the slope of the linear growth of the GE gives 
the KS entropy.

\subsection{Final temperature of the fragments}

Since the average density rapidly changes from the low initial
values to almost the saturation during the same interval of time,
we conclude that the temperature of the fragments increases with respect
to the initial value (T=3 MeV). This is indeed the case, as it is 
shown in fig.7(a).
There we plot the one-body distribution function, averaged over 
a large number of cells in coordinate space, as a function of the
energy. The histogram indicate the results of the numerical calculation
(one typical event), 
whereas the solid line is a fit with a Fermi-Dirac distribution function
at temperature $T= 7~MeV$. This finding is weakly dependent on the 
initial average density, as clearly shown in fig.8(a), where
the final fragment temperature is slightly higher (T=8 MeV) and the
system is initialized at density $\bar \rho/\rho_0 = 0.3$.
Some details on the study of the distribution function are given
in Appendix.

In figs.7(b) and 8(b) we display a typical distribution function 
of the gas vs. the energy density, averaged over a finite number of cells.
From this distribution, which does not resemble a maxwellian because of the 
high momenta tails, we can extract the temperature of the gas
(see Appendix) and this turns out to be much larger than the fragments'
temperature. More precisely we get for the initial density 
$\bar \rho/\rho_0 = 0.55 $~~  $T_{gas}= 30.4~ MeV$, 
while  $T_{gas}= 15.8~ MeV$ for $\bar \rho/\rho_0 = 0.3 $. Therefore
the system formed by fragments plus gas is on the whole not
equilibrated, being the two components at very different final temperatures.
Though these numbers are not the average over many simulations, 
this result is typical, {\it i.e.} repeating the calculation one 
gets a similar result.

This fact is consistent with some features observed in the experimental
data. In ref.\cite{indra} it has been found  that the apparent
temperatures extracted from the slopes of the spectra of the 
emitted light particles
(p,d,$^3He$) are in general greater than those measured by other 
experimental groups \cite{gsi,eos2} through the ratio of the double
isotope yields \cite{albergo}. In ref. \cite{eos2} it has been claimed that 
such particles are emitted in a first stage of the collision and not from 
a source in thermal equilibrium at variance with the intermediate mass 
fragments (IMF). Then, notwithstanding the simplifications of our 
numerical simulations, 
we find close analogies with the experimental data. 
At the moment, the fact that   
we obtain  a gas with a temperature higher by a factor of two-three cannot 
allow definite conclusions. The comparison with the experimental data 
is more qualitative than 
quantitative.  However, both theory and experiments seem to strongly indicate
 that the temperatures
extracted from the light particles might  not carry information on the 
thermal equilibrated source of MF. Thus 
the apparent temperatures extracted from the
slopes may  be misleading for extracting the nuclear caloric curve and one 
should be very careful in using them.   

Finally, in fig.9 we draw the 2D equation of state (EOS) for nuclear matter
with the same Skyrme forces employed for this model, see refs.\cite{bbr2} 
for more details. 
The solid line is the isothermal curve at temperature T=0,
the dashed line at T=3 MeV and the patterned area represents the region 
where our fragments lie after the dynamical evolution started in the 
spinodal region. The circle (square) 
represents the final state of the fragments 
when the system is 
started at $\bar \rho/\rho_0 = 0.3 ~(0.55)$.
Error bars indicate uncertainties on their 
final density.
We see that the final formed fragments are stable and close to 
thermodynamical equilibrium. This plot confirms that the numerical
simulation is fully consistent and reliable. From this result and from the 
moments of the  mass distributions reported in  ref.\cite{bbr2},
 we can argue that while the events 
corresponding    to an initial density 0.55 are very close to the 
MF phase transition point  (power law in the mass distribution), 
those corresponding to 0.3 correspond to a fragment production  of 
smaller size  and therefore to a higher initial excitation energy. 
The value we get for the temperature of the fragments in the two cases 
(7 and 8 MeV respectively) 
therefore indicate a rise in the caloric curve of the kind observed by the
Aladin and the Indra groups  \cite{gsi,indra}. Again the comparison is more
qualitative than quantitative. 

\section{Conclusions}

Within the framework of the Vlasov equation solved in a 2D lattice by means 
of a deterministic algorithm we have studied the process of fragment 
formation in MF events. The model has 
already been used with success in the past to clarify the dynamics of nuclear 
matter  inside the spinodal region.  In the present paper we have studied 
the time evolution of the coarse-grained GE when MF occurs.
It has to be stressed that two-body collisions are missing in our model.
Despite that,
we have found that the GE increases rapidly at the 
moment of fragment formation and
saturates soon after. This behavior confirms the role of chaos in filling 
dynamically the available phase space ( in the coarse-grained sense ), 
as measured by the GE increase.
In fact, only if the dynamics is chaotic,
and therefore mixing, the initial smooth distribution is able to spread
in phase space until the reaching of equilibrium. In this process
the initial distribution is expected to change shape and form a rather 
irregular pattern with a sharp
increase of the size of its boundary. This means that 
the momentum
distribution will be strongly dependent on the position. Consequently
the density will tend to vary from one place to another,
which favours fragment formation.
This picture is consistent with the fact that the main source of
entropy turns out to be the surface regions of the system.
These results are at variance with those obtained using the linear response 
which is not able to thermalize the system and 
does not give any growth of entropy.

The general trend of the entropy values are actually not far from the ones
discussed in the literature in connections with the analysis of
several experimental data \cite{kap,fopi}. Of course, due to
the schematic character of the model, no detailed comparison with
experiments is possible.

We have found
a relaxation to equilibrium for what 
concerns intermediate mass fragments (IMF). The density of
the final fragments is
fully consistent with the calculated EOS. 
On the other hand we find a gas 
which is not in equilibrium with the liquid part (IMF), having a temperature 
which is a factor of 2-3 higher. Though the role of the collision term 
should be better investigated in this respect, this feature has been found also 
experimentally and could be very general. Finally the temperature of the 
fragments corresponding to higher excitation energy is slightly larger.
The latter is a very preliminary result which will be investigated 
with more detail in the future. 

\section{Acknowledgments}

 We thank V. Latora for helpful and stimulating discussions. 
 Clarifying discussions with M. Colonna are also  acknowledged. 


\appendix

\section{}

In order to investigate equilibrium properties of nuclear matter 
we studied the momentum distribution function $f_{bulk}(p)$

\begin{equation}
f_{bulk}(p)~ =~ \frac{ \int f(x,y,p_{x},p_{y})~ 
\delta (x - x_{F})~ \delta (p - P^{1/2})~ d \Gamma }
{\int \delta (x - x_{F})~ \delta (p - P^{1/2})~ 
d \Gamma }
\end{equation}

being $P = p_x^2 + p_y^2$, $f(x, y, p_x, p_y)$ is the occupation probability, 
$d \Gamma$ is the phase space volume element and 
$x_F$ is an ensemble of cells in coordinate space, inside a given fragment, 
over which the distribution function is averaged. 

In order to calculate the fragment temperature we have minimized the
function

\begin{equation}
\chi~^{2}(\mu , T) = \frac{\int | f_{bulk} (\epsilon)~ -  
~f^{FD} (\epsilon , \mu , T) |^{2} 
d \epsilon }{\int d \epsilon }
\end{equation}

being $\epsilon =  p^{2} / 2 m$ and  
$f^{FD} (\epsilon , \mu , T)$ is the Fermi-Dirac distribution

\begin{equation}
f^{FD}(\epsilon , \mu , T) = \frac{ 1 }{ 1 + exp(( \epsilon - \mu )/ T)}
\end{equation}

keeping $\mu$ e $T$ as free parameters to be fitted. 

We proceed in a similar way in order
to investigate the equilibrium properties of the gas
($\bar \rho \leq 0.05$). For this purpose we have studied the distribution 
function $f_{gas}(p)$

\begin{equation}
f_{gas} (p)~ =~ \frac{ \int f(x,y,p_{x},p_{y})~ 
\delta (x - x_{g})~ \delta (p - P^{1/2})~ d \Gamma }
{\int \delta (x - x_{g})~ \delta (p - P^{1/2})~ 
d \Gamma }
\end{equation}

being $f$ the occupation probability and 
$d \Gamma$ the phase 
space volume element. $x_g$ is an ensemble of gas cells 
in coordinate space over which the distribution function is averaged. 
Therefore we have calculated the effective temperature of the gas as

\begin{equation}
T_{eff} = \frac{\int d\Gamma \frac{p^2}{2m} f_{gas}(p)}
{\int d\Gamma f_{gas}(p)}
\end{equation}
\noindent
In fact considering that we are in 2D and in classical mechanics 
- due to the high temperatures - the average of the kinetic energy  is
equal to the temperature.
In the case of the discretized version of the Vlasov equation,
instead of the integrations and of the $\delta$-functions,
the appropriate summations and discrete $\delta$'s must be used.


\bigskip

$(i)$   E-mail: atalmi@ct.infn.it 

$(ii)$  E-mail: baldo@ct.infn.it 

$(iii)$ E-mail: burgio@ct.infn.it 

$(iv)$  E-mail: rapisarda@ct.infn.it


\newpage 
\begin{figure}
\caption[]{\footnotesize
{Filling of phase space for the standard map (top panels) for two values 
of K=0.05,2 
and respective time evolution of the coarse-grained GE. The slope of the 
linear rise of S gives the KS entropy, i.e. the largest 
Lyapunov exponent in this case \cite{baranger}.}}
\end{figure}
\begin{figure}
\vskip 1cm
\caption[]{\footnotesize
{For a typical event, we plot the space density $\bar \rho(x)$ 
at different times. The initial average density 
$\bar \rho$ is equal to 0.55$\rho_0$. Please notice the different scale used
in the top panel.}}
\end{figure}
\begin{figure}
\vskip 1cm
\caption[] {\footnotesize {
A typical 
 time evolution of the occupation number is shown in panels (a) and (c) 
for different choices of cell size $\Delta x$ and the gaussian width
$\mu$. In panels (b) and (d) the 
total entropy per particle is shown. The solid lines represent calculations
performed with $\Delta x = 0.66 fm$, whereas the dashed line the ones
at $\Delta x = 0.33 fm$. See text for details.}}
\end{figure}
\begin{figure}
\vskip 1cm
\caption[]{\footnotesize 
{The time evolution of $\alpha$ (the number of coordinate space cells) 
 and $\sigma$ are shown in panels
(a) and (b) for the bulk (circles) and the surface (dashed line). 
The gas components,
{\it  $\alpha_{gas}$ and $\sigma_{gas}$} (dotted line) are shown in panels 
(c) and (d) for 
a typical trajectory at an initial density $\bar \rho/\rho_0 = 0.55$.
The solid line in panels (b) and (d) represents the total entropy 
per particle. Arrows indicate the fragmentation time $\tau_{frag}$.}}
\end{figure}
\begin{figure}
\vskip 0.7cm
\caption[] {\footnotesize {
Same as Fig.4, but for 
a typical trajectory at an initial density $\bar \rho/\rho_0 = 0.3$.}}
\end{figure}
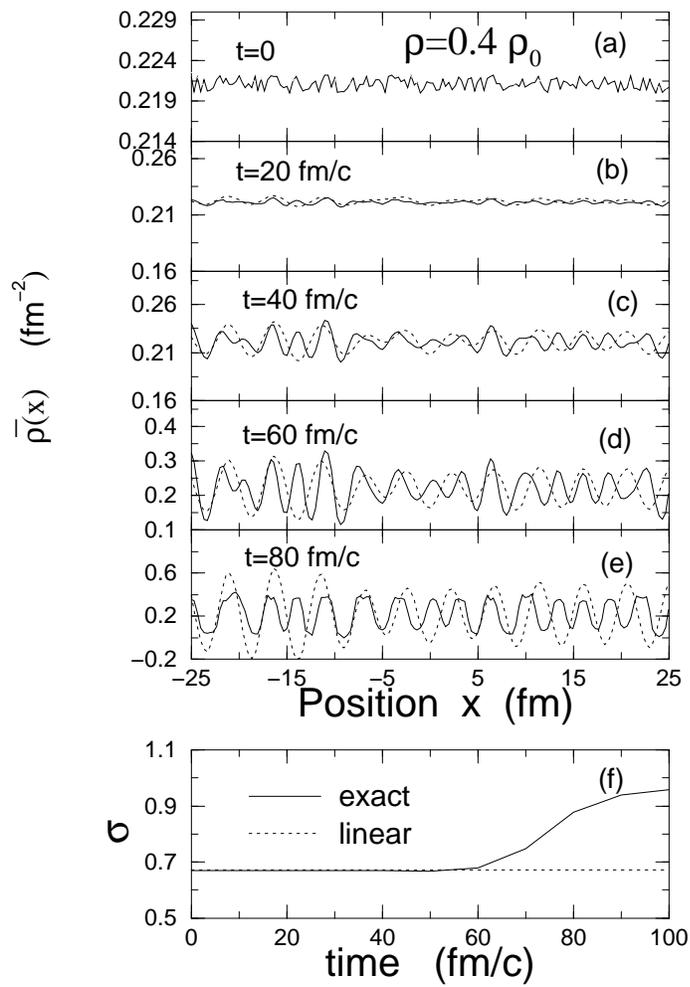
\begin{figure}
\caption[] {\footnotesize {
The exact numerical evolution (full line) and the linear response one
(dashed line) 
are compared for the density profiles (panels (a)-(e)) 
and for the entropy evolution (panel (f)).
The linear response is able to follow the exact
evolution approximately only up to 50 fm/c. For later times,
the difference between the two profile increases and the evolution of the 
linear response becomes unreliable: the violation of energy and particles
number is substantial already at 60 fm/c, while  at 80 fm/c  the profile
becomes also negative.
However the entropy production is the real 
big difference between the two approaches. As shown in panel (f) the linear 
response does not produce any entropy growth
at variance with the exact simulation. See text for further details.
}}
\end{figure}

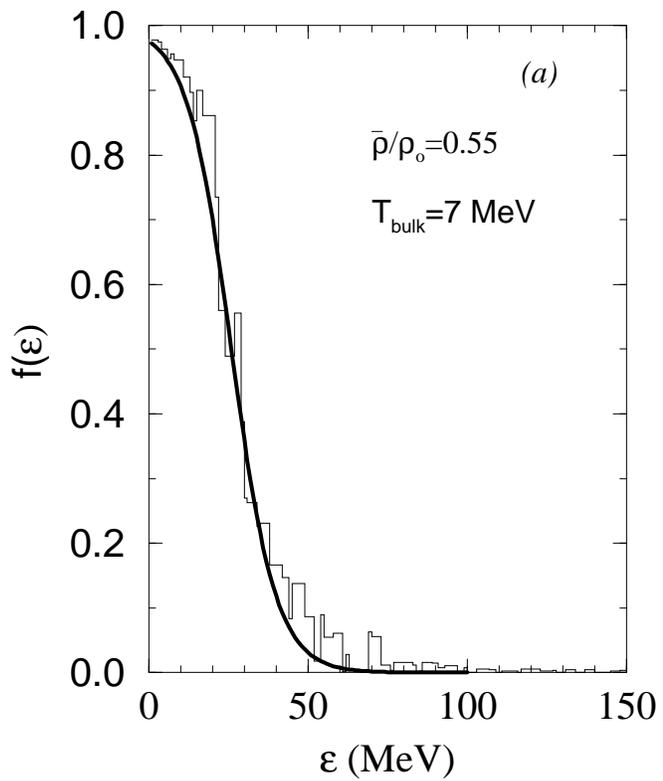
\begin{figure}
\caption[] {\footnotesize {
In panel (a)   we plot 
(histogram) a typical
 one-body distribution function  
$f(\epsilon)$, $\epsilon = P^2 /2m$, calculated in the bulk 
of the fragments (and averaged over all fragments) 
at the final time t=200 fm/c. 
The initial average density is
$\bar \rho / \rho_0 = 0.55$. The solid line represents
the fit performed with a Fermi-Dirac distribution with a temperature  
$T$ (see Appendix). We get an average fragment temperature T=7 MeV. 
In panel (b) we show the distribution function for the gas.
In this case the calculation of the temperature as discussed in the Appendix
gives T=30.4 MeV.
}}
\end{figure}
\begin{figure}
\caption[] {\footnotesize {
Same as Fig.7 for $\bar \rho / \rho_{0} = 0.3$ at the final
time t=100 fm/c.}} 
\end{figure}
\begin{figure}
\vskip 1cm
\caption[] {\footnotesize {
The pressure is plotted vs. the density for a 2D system with the same
Skyrme forces used for the model investigated (see ref.\cite{bbr2}).
The solid line represents the EOS at zero temperature, the dashed line the one at
T=3 MeV, whereas the hatched area encloses the EOS's between T=6.5 MeV
and T=8.5 MeV. The filled circle represent the average final state of the 
fragments  when the nuclear system is initialized 
at $\bar \rho/\rho_0 = 0.3$, whereas the filled square 
concerns the one at $\bar \rho/\rho_0 = 0.55$. 
The error bars indicate  uncertainties on their final density. }}
\end{figure}
\vfill
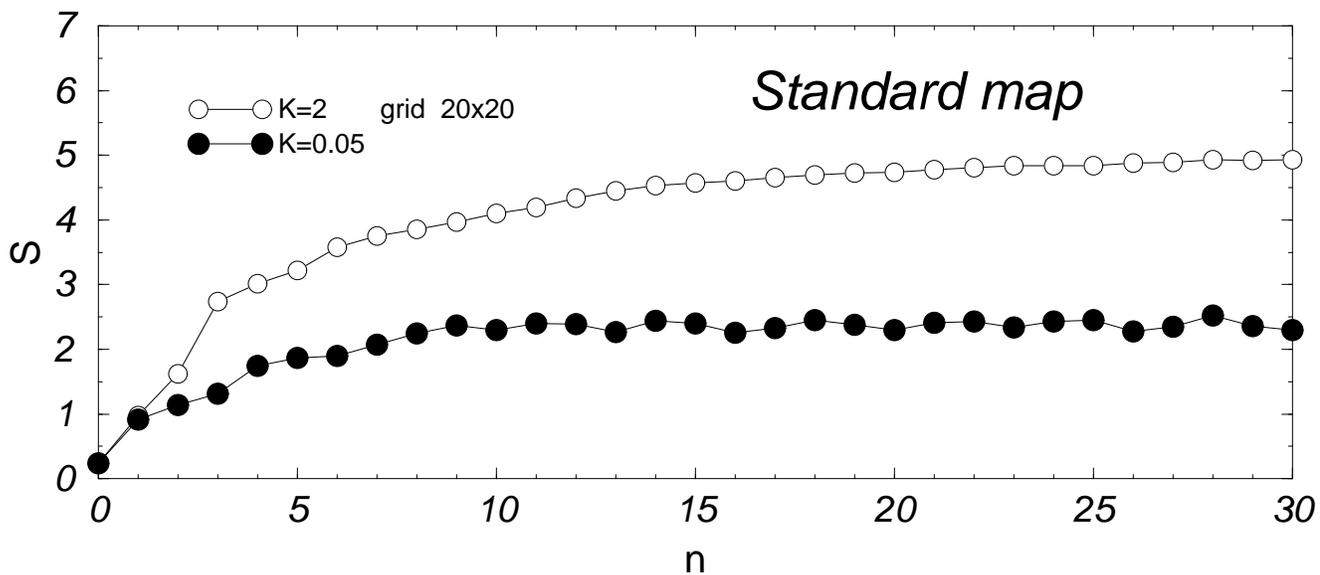
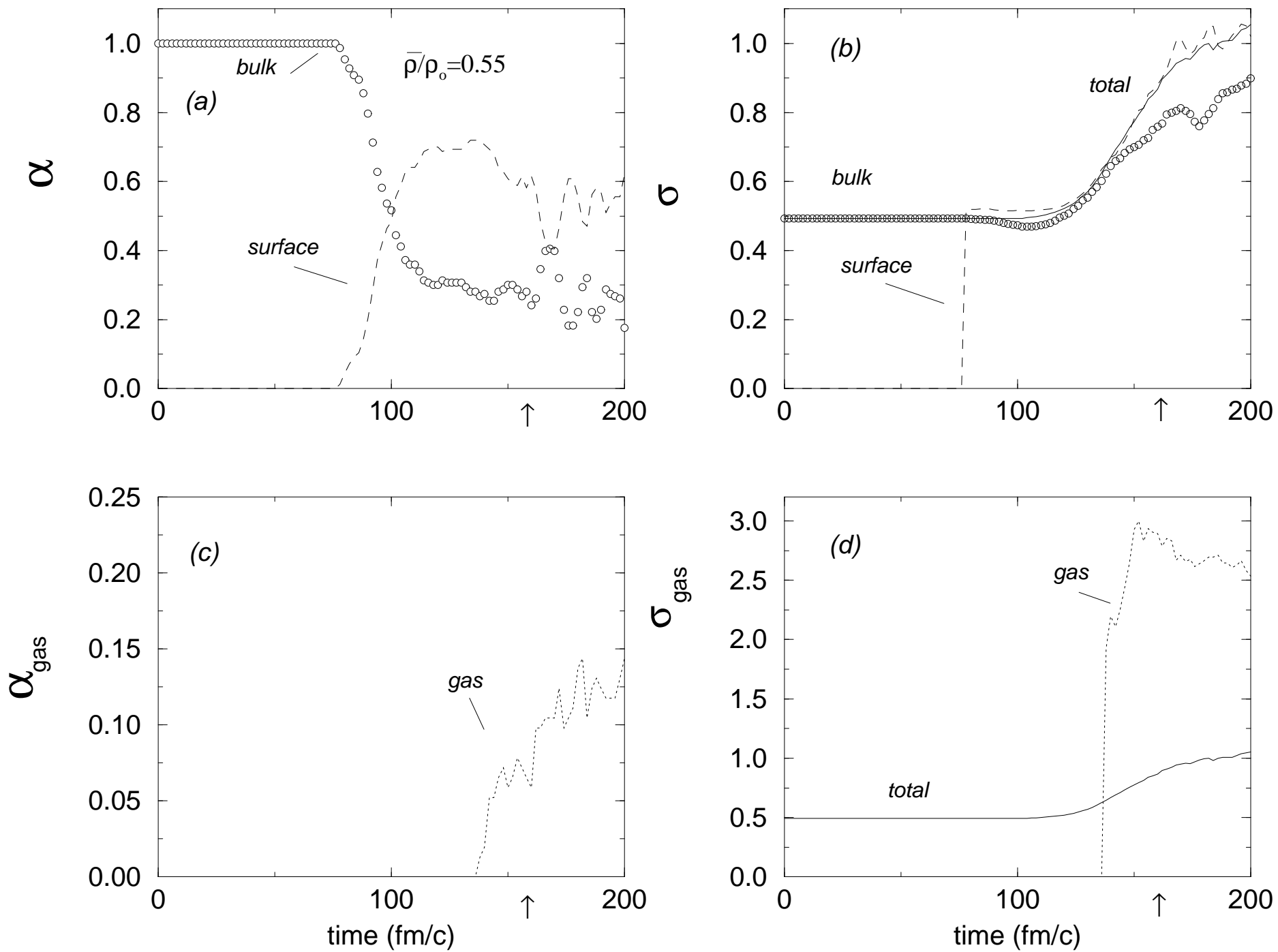
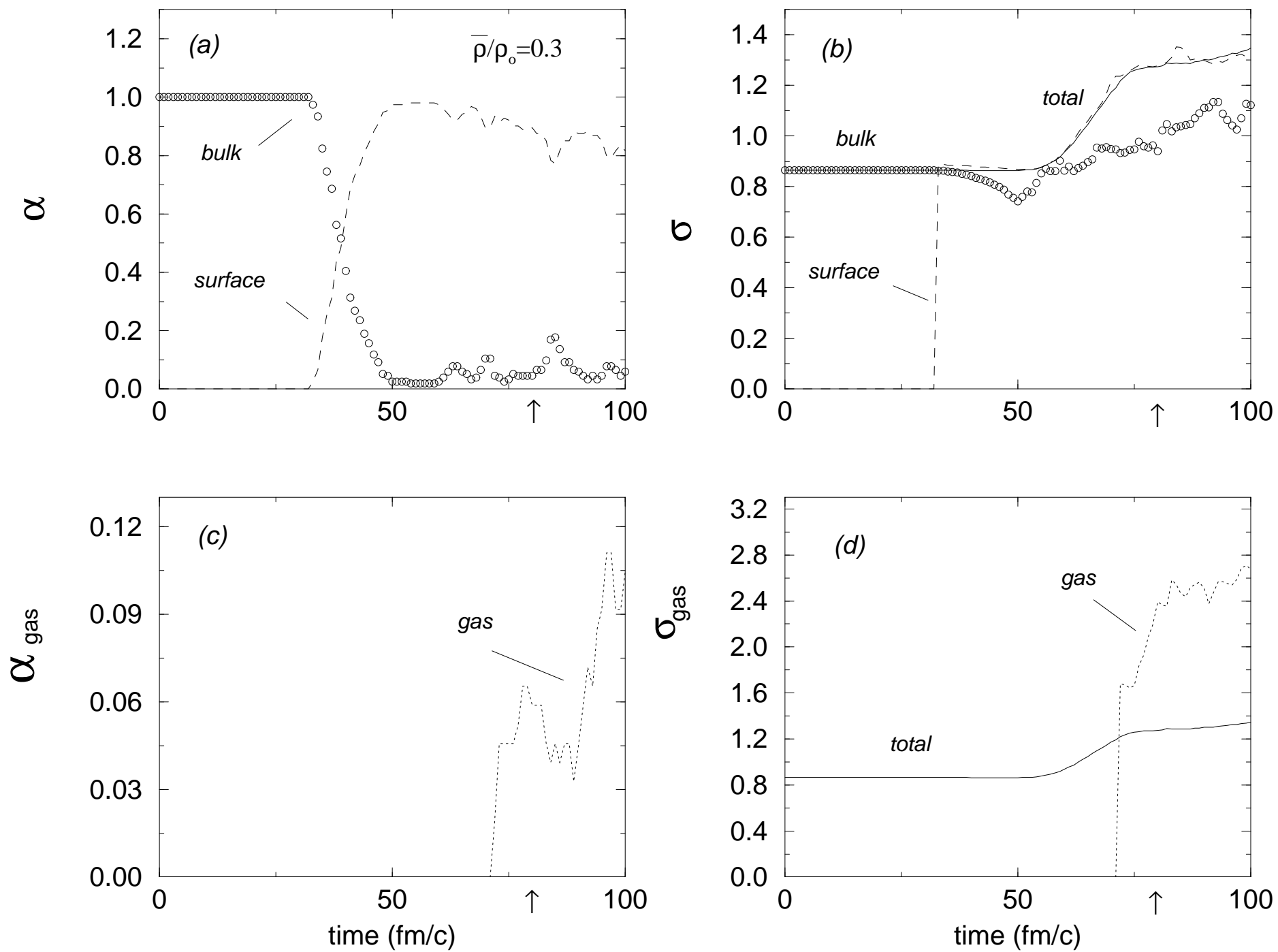
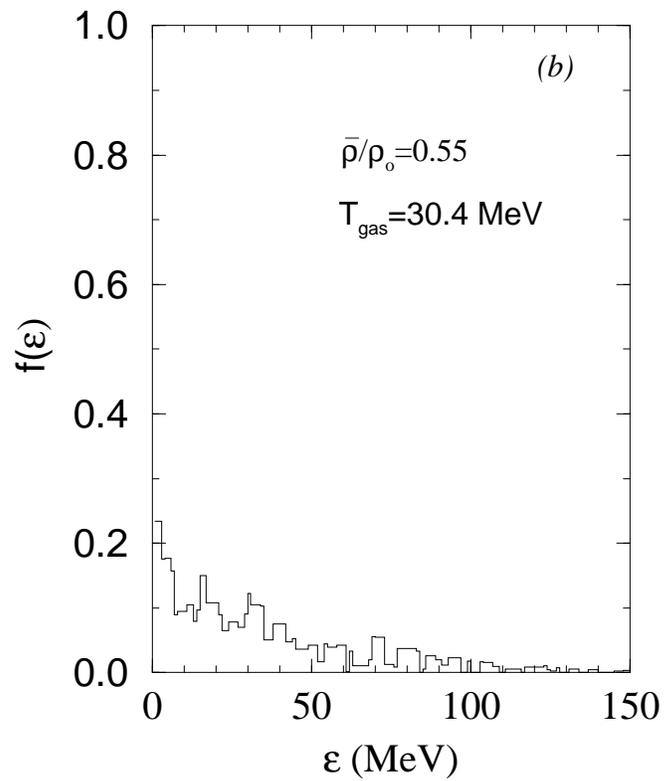
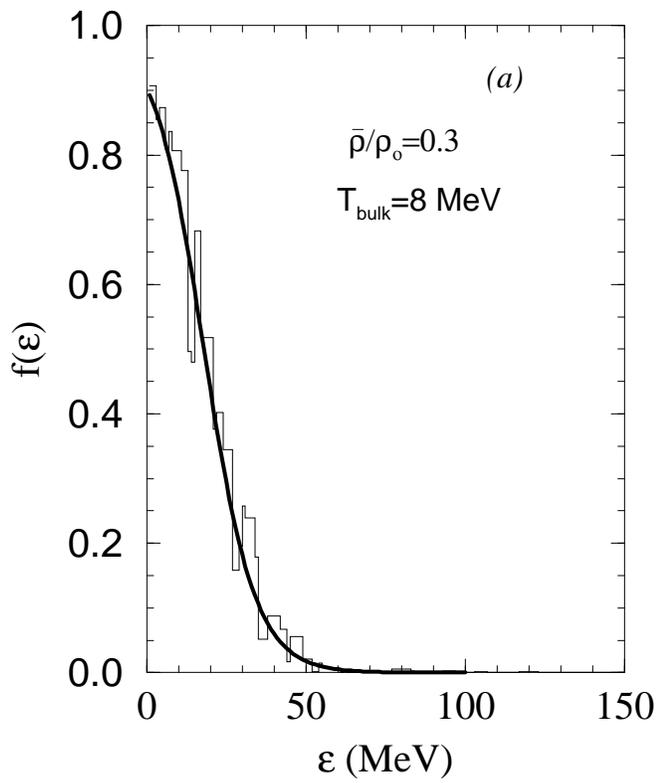
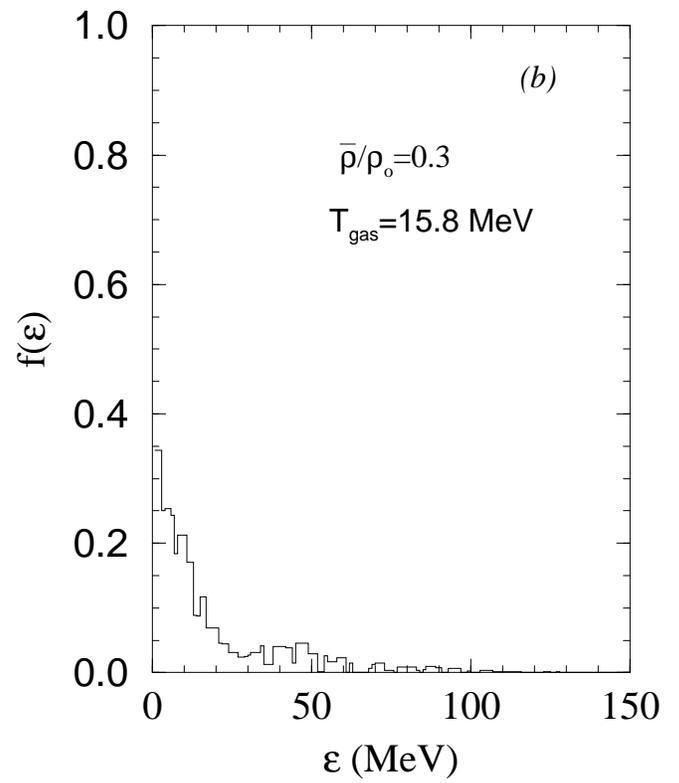
\end{document}